\documentclass [a4paper,11pt]{article}
\usepackage[english]{babel}
\usepackage{graphicx,amsmath,amsfonts}
\usepackage{color}

\renewcommand{\a}{\alpha}
\renewcommand{\b}{\beta}
\renewcommand{\d}{\delta}

\renewcommand{\ss}{\sigma}

\newcommand{\du}{\partial_{1}}

\newcommand{\be}{\begin{equation}}
\newcommand{\ee}{\end{equation}} 
\newcommand{\eei}{\end{equation}\indent\indent}
\newcommand{\bc}{\begin{center}}
\newcommand{\ec}{\end{center}}
\newcommand{\ber}{\begin{eqnarray}}
\newcommand{\ear}{\end{eqnarray}}
\newcommand{\ba}{\begin{array}}
\newcommand{\ea}{\end{array}}

\def\case#1/#2{\textstyle\frac{#1}{#2} }

\begin{document}

\begin{center}
$\Large{\textbf{General relativistic corrections and non-Gaussianity}}$\\
$\Large{\textbf{in large-scale structure}}$
\\[1cm]

$\large{ \text{Eleonora Villa}^{a,b} ,\text{Licia Verde}^{c,d} \,\text{and Sabino Matarrese}^{e,f}}$
\\[0.5cm]
\small{
\textit{$^{\rm a}$ Institute of Cosmology and Gravitation, University of Portsmouth,
Dennis Sciama Building, Burnaby Road, PO1 3FX Portsmouth, United Kingdom}}
\\[1mm]
\small{
\textit{$^{\rm b}$ Dipartimento di Fisica, Universit\`a degli Studi di Milano, via Celoria 16, 20154 Milano, Italy}}
\\[1mm]
\small{
\textit{$^{\rm c}$ 
ICREA (Instituci\'o Catalana de Recerca i Estudis Avan\c{c}at) and ICC (Institut de Ciencies del Cosmos),
Universitat de Barcelona (UB-IEEC), Marti i Franques 1, Barcelona 08028, Spain }}
\\[1mm]
\small{
\textit{$^{\rm d}$ 
Institute of Theoretical Astrophysics, University of Oslo, 0315 Oslo, Norway}}
\\[1mm]
\small{
\textit{$^{\rm e}$ Dipartimento di Fisica e Astronomia ``G. Galilei", Universit\`a degli Studi di Padova and INFN Sezione di Padova, via Marzolo 8, 35131 Padova, Italy}}
\\[1mm]
\small{
\textit{$^{\rm f}$ Gran Sasso Science Institute, INFN, viale F. Crispi 7, 67100 L'Aquila, Italy}}


\end{center}

\vspace{2cm}

\begin{abstract}
General relativistic cosmology cannot be reduced to linear relativistic perturbations superposed on an isotropic and homogeneous (Friedmann-Robertson-Walker) background, even though such a simple scheme has been 
successfully applied to analyse a large variety of phenomena (such as Cosmic Microwave Background primary anisotropies, matter clustering on large scales, weak gravitational lensing, etc.). 
The general idea of going beyond this simple paradigm is what characterises most of the efforts made in recent years: the study of second and higher-order cosmological perturbations including all general relativistic contributions --
also in connection with primordial non-Gaussianities -- the idea of defining large-scale structure observables directly from a general relativistic perspective, the various attempts to go beyond the Newtonian approximation 
in the study of non-linear gravitational dynamics, by using e.g., Post-Newtonian treatments, are all examples of this general trend. Here we summarise some of these directions of investigation, 
with the aim of emphasising future prospects in this area of cosmology, both from a theoretical and observational point of view. 
\end{abstract}

\newpage
\tableofcontents
\vspace{.5cm}

\newpage

\section{Introduction}
The large-scale structure (LSS) of the Universe is the result of the evolution of cosmological perturbations generated during inflation: quantum fluctuations of the inflaton field set the seeds of curvature perturbations at primordial epochs. 
The theoretical study of structure formation connects the early, quasi-homogeneous Universe with the highly inhomogeneous one observed today. Gravitational instability is the process that drives the growth of cosmological perturbations into cosmological structures and is governed by the equations provided by General Relativity (GR). However,  given the large dynamic range required to study the formation of LSS, more than six orders of magnitude in density and five or more orders of magnitude in scale, any such study is carried out with the use of different approximations, depending on the specific range of applicability. We study large and small scales in two different ways. 
Relativistic perturbation theory around a homogeneous and isotropic background, the Friedman-Robertson-Walker (FRW) solution, is used at large scales (of the order of the Hubble horizon), where the growth of structures is at an early stage. Even if second (or even higher)-order perturbations are considered, e.g., to compute non-Gaussianity (NG), the matter density perturbations must be small for this description to apply. At smaller scales, well inside the Hubble horizon, GR is replaced by Newtonian gravity and non-linear gravitational instabilities are studied by means of Newtonian N-body simulations or by approximate analytical expressions obtained for example via perturbation theory,  the Zel'dovich approximation, \cite{zel}, or its extensions. 
The shortcoming of this treatment is, obviously, that it does not include GR effects.
This split (early-times and large scales vs late times and small scales)  simplifies things immensely and in these two limits the approximations adopted do not introduce any significant limitation. 
Away from these two limits however the approximations are not guaranteed to hold true.

In particular, there are situations where we have to overtake this distinction, e.g., if we want to study structure formation including relativistic effects in a context where non-linearities are important, even within the Hubble horizon. Moreover upcoming large-scale structure surveys (such as  the Euclid \cite{euclid} galaxy survey) will probe scales approaching the Hubble horizon, where the Newtonian approximation is no longer valid. 
Finally, we have to take into account that the observations are performed along our past light cone, not at a fixed instant of the cosmic time: on large scales and at large distances it becomes necessary to include light-cone and gauge relativistic effects for the correct interpretation of all the data (for a pioneering analysis of the relevance of relativistic cosmology for LSS, see \cite{ellis71}).

The outstanding accuracy reached or reachable by the observations, motivates the theoretical efforts to obtain accurate predictions of these effects. 

A relativistic treatment of  galaxy clustering is essential to understand the relation between the cosmological dynamics in GR (or in alternative theories for gravitation that can be tested) and the observables: cosmological perturbations in the space-time metric affect the photon path via lensing and redshift-space distortion, frame dragging, gravitational redshift, Sachs-Wolfe effect etc. For a recent assessment of the relevant effects on the  galaxy correlation function at very large scales see \cite{Roy}.  A consistent derivation of all these effects at large scales was given at first order in \cite{Yoo2009, jeong}. Recently these analyses were extended at second order, see \cite{Yoo2014} and \cite{veneziano1, didio2014, bertacca14res} for 
the calculation of some cosmological observables. A second-order calculation is crucial because the linear (scalar) modes generate non-linear (vector and tensor) ones, many of which are not taken into account in Newtonian theory, and play an important role in the dynamics of perturbations.

A proper application of the GR formalism to observations would however also require a GR treatment for galaxy bias which is highly non-trivial, because of the gauge choice, which is related to the choice of the time slicing for the matter density, see in particular the discussion in \cite{wands&slosar}-\cite{brunibias}. \\

In this context, a closely related issue is that of the study of deviations from Gaussian statistics of the cosmological perturbation field. In the simplest (single-field, slow roll) inflationary models  the  deviations from Gaussianity of the initial fluctuation field are small and unobservable. Non-linear gravitational evolution induce specific non-Gaussian signatures which are of high-signal to noise. In addition, since inflation generates perturbations in the potential but large-scale structure observations map the density perturbation, on scales comparable to the horizon GR effects might introduce a mis-match of the statistical properties of the two fluctuation fields.

Future, high-precision measurements of the statistics of the density and peculiar velocity fields  from the LSS will allow us to pin down the non-Gaussian signal, thus providing a tool complementary to studies of the cosmic microwave background  (CMB) anisotropies. 
It is therefore important to interpret the non-Gaussian signal in the LSS, understand how it can be separated out from the non-linear one and investigate if there are clean GR contributions.

The plan of the paper is as follows. In section 2 we consider sufficiently large scales and present the basic guidelines to capture the relation between primordial non-Gaussianity, 
dark matter density and gravitational potential, at second order in perturbation theory. 
We then consider the GR correction to  the large-scale halo bias in Sec. 3. In section 4 we consider the post Newtonian  approximation of GR in a fully non-perturbative perspective, which can be used for a unified description for linear and non-linear scales, including the first GR corrections. All our calculations are performed assuming a flat universe with irrotational and pressure-less cold dark matter plus a cosmological constant as FRW background.  The results are presented in the synchronous (and time-orthogonal) and comoving gauge which is appropriate for LSS studies (e.g., to estimate the Lagrangian bias of dark matter halos) as pointed out in \cite{wands&slosar}. Finally we conclude in section \ref{sec:conclusions}.

\section{GR effects on large scales} \label{NGlarge}

Non-Gaussianity (NG) has become a new branch in early Universe cosmology. If detected, primordial NG would be the most informative fingerprint of the origin of structure in the Universe, 
probing physics at extremely high energy scales, including particle physics interactions (for recent reviews see \cite{reviewNG, Liguorietal, Chen}), as well as possible deviation from GR at such scales \cite{Bartolo:2014kaa}. 
Constraints on the amplitude and form of primordial NG allow one to discriminate among competing mechanisms for the generation of the cosmological perturbations in the early Universe, in that
different inflationary models predict different amplitudes, shapes, and scale dependence of NG. 

A convenient way to describe primordial non-Gaussianity is to consider the curvature perturbation of uniform density hypersurfaces. This is a gauge-invariant quantity which remains constant on super-horizon scales after 
it has been generated during a primordial epoch (and possible isocurvature perturbations are no longer present) and therefore provides the initial conditions.
In general, we may  parameterise the primordial NG level in terms of the conserved curvature perturbation, which up to second order is expanded as, see e.g., \cite{Bartolo:2005fp}, 
$\zeta =  \zeta^{(1)} + (1/2)\,\zeta^{(2)} =\zeta^{(1)} + (a_{\rm nl} -1) \star (\zeta^{(1)})^2 + \cdots$, where $\star$ stands for a convolution, as the parameter $a_{\rm nl}$ for different inflationary scenarios may depend on scale and configuration\footnote{For constant $a_{\rm nl}$ (or $f_{\rm NL}$, see below), i.e., for to so-called  local NG, the convolution reduces to a simple multiplication.}, for a review see \cite{reviewNG}.
In the standard single-field inflation $\zeta^{(2)}$ is generated during inflation and is proportional to $\left( \zeta^{(1)}\right)^2$, the constant of proportionality $(a_{\rm nl} -1)$ is ${\cal O}\left(\epsilon,\eta\right)$
 where $\epsilon$ and $\eta$ are the usual slow-roll parameters. 

When focusing on observations of the CMB on large angular scales, the contribution of primordial NG is transferred linearly to leading order; 
therefore it is convenient to introduce the potential $\Phi$, directly proportional to the comoving curvature perturbation, as $\Phi=3/5\zeta$; in matter domination and on super-horizon scales, $\Phi$ is equivalent to Bardeen's gauge-invariant gravitational potential. However, when considering LSS the natural quantity to consider is the matter density contrast $\delta$, which takes NG contributions from various sources, so that its relation to $\zeta$ is not as simple. 
It is therefore not straightforward to single-out the primordial NG signal in the matter density contrast and compare it directly with CMB constraints. Here we address this by ``meeting half way", i.e. by connecting $\zeta$ and $\delta$ to the gravitational potential $\Phi$. Other approaches have also been explored (e.g. \cite{wands&slosar, BHMW14}) leading to similar conclusions.

To characterise the amplitude of NG in the matter density contrast beyond the usual second-order Newtonian contributions we follow \cite{BMPR2010} and introduce an effective gravitational potential $\Phi$ obeying the standard
Poisson equation 
\begin{equation}
\label{P}
-\nabla^2 \Phi=\frac{3}{2}\Omega_m  {\cal H}^2 \delta, 
\end{equation}
and at an initial epoch, deep in matter domination, we write  
\begin{equation}
\label{fnlphiin}
\Phi_{\rm in}=\Phi^{(1)}_{\rm in}+f_{\rm NL}\star(\Phi^{(1)2}_{\rm in}-\langle \Phi^{(1)2}_{\rm in} \rangle).
\end{equation}
In these expressions $\Omega_m(\tau)$ is the FRW matter density parameter at conformal time $\tau$,  $\Phi \propto g(\tau)$, $g(\tau)$ being the ratio of the linear 
Newtonian growing mode and the FRW scale factor $a$, and ${\cal H}$ denotes $a^{-1}da/d\tau $. 
The dimensionless non-linearity parameter $f_{\rm NL}$ sets the level of quadratic NG and is also defined to make contact with the primordial NG: 
because of the linear relation between $\zeta$ and $\Phi$, in general $f_{\rm NL}=5/3(a_{\rm nl} -1)$. A Gaussian distribution of primordial perturbations corresponds to $f_{\rm NL}=0$, $(a_{\rm nl} =1)$. 
Note  that the bispectrum corresponding to Eq.~(\ref{fnlphiin}) is 
\begin{equation}
B_{\Phi}(k_1,k_2,k)=2 f_{\rm NL} P_{\Phi}(k_1) P_{\Phi}(k_2) +cyc.
\label{eq:Bfnl}
\end{equation}
While this is strictly correct for  constant $f_{\rm NL}$ (local non-Gaussianity) we can introduce a scale- and shape-dependent 
non-linearity parameter, $f_{\rm NL}(k_1,k_2,k)$, which generalises the standard definition of \cite{verde} inferred from the Newtonian gravitational potential. In this case Eq.~(\ref{eq:Bfnl}) would involve a convolution rather than a simple multiplication.

All NG in the matter density contrast can be written explicitly by  expressing the 
Fourier space $\delta_{\bf k}$, in 
terms of the linear density contrast, $\delta^{(1)}_{\bf k}$, and  
defining the kernel ${\cal K}_\delta({\bf k}_1,{\bf k}_2;\tau)$  as
\begin{eqnarray}
\label{defkerneldensity}
\delta_{{\bf k}} (\tau)&=& \delta^{(1)}_{{\bf k}} (\tau)+\frac{1}{2} \delta^{(2)}_{{\bf k}}(\tau)
\\
&= & \delta^{(1)}_{{\bf k}} (\tau)+\int \frac{d^3{\bf k}_1d^3{\bf k}_2}{(2 \pi)^3} \,{\cal K}_{\delta}({\bf k}_1,{\bf k}_2;\tau) \delta^{(1)}_{{\bf k}_1}(\tau) \delta^{(1)}_{{\bf k}_2}(\tau)
\delta_D ({\bf k}_{1}+{\bf k}_{2}-{\bf k})\, .
\end{eqnarray}
We can  further write the kernel as 
\begin{eqnarray}
\label{kddthth}
{\cal K}_{\delta}({\bf k}_1,{\bf k}_2;\tau)  &=&  {\cal K}_{\delta}^N({\bf k}_1,{\bf k}_2;\tau)+\frac{3}{2} \Omega_m {\cal H}^2 f_{\rm NL}({\bf k}_1,{\bf k}_2,\tau) \frac{g_{\rm in}}{g(\tau)} 
\frac{k^2}{k_1^2 k_2^2}\, ,
\end{eqnarray}
where $k^2 \equiv |{\bf k}_1 + {\bf k}_2|^2$ and ${\cal K}_{\delta}^N({\bf k}_1,{\bf k}_2;\tau)$ is the second-order Newtonian kernel  arising from non-linear gravitational evolution.
 
 To find an explicit expression for $f_{\rm NL}$, consider the GR density contrast up to second order
\begin{eqnarray}\label{deltaS}
\delta & =&  \frac{100}{9{\cal H}_0^2} \left[f(\Omega_{0m}) + \frac{3}{2} \Omega_{0m}\right]^{-1} \bigg\{ D_+(\tau)\left[\left(\frac{3}{4} 
- a_{\rm nl}\right)\left(\nabla \varphi_{\rm in}\right)^2 
+ (2 - a_{\rm nl} ) \varphi_{\rm in} \nabla^2 \varphi_{\rm in} \right]+
\nonumber  \\
 & +&D_+(\tau)\frac{3}{20}\nabla \varphi_{\rm in}+ \frac{D_+^2(\tau)}{14 {\cal H}^2_0} \left[f(\Omega_{0m}) + \frac{3}{2} \Omega_{0m}\right]^{-1} \left[ 5 \left( \nabla^2 \varphi_{\rm in} \right)^2 +
2 \partial^i \partial^j \varphi_{\rm in}  \partial_i \partial_j \varphi_{\rm in} \right]  \bigg\}\,,
\end{eqnarray}
were $D_{+}$ denotes the linear growing mode, $f(\Omega_m) \equiv d \ln D_+ / d \ln a$; the subscript $0$ denotes the quantity at present time, while the subscript ${\rm in}$ refers to quantities at the initial time.  
This shows how the information on the primordial NG, set on super-Hubble scales, flows into smaller scales, once the mode re-enters the horizon: the Newtonian first and second-order terms in the second line are insensitive to the non-linearities in the initial conditions, whereas it is the second-order PN contribution of the first line which carries the relevant information on primordial NG (as was first pointed out in \cite{BMR2005}) \footnote{While the PN contribution is exact, the Newtonian expression 
is valid under the approximate that $f(\Omega_m)/\Omega_m \approx 1$. See \cite{BHMW14} for the exact solution.}. Replacing the last expression in Eq.~\eqref{defkerneldensity}, allows us to re-express this result in terms of $\Phi$. 
This yields an explicit expression for $f_{\rm NL}$,
\begin{equation}
f_{\rm NL}({\bf k}_1,{\bf k}_2)  =   \left[ \frac{5}{3}(a_{\rm nl}-1) + f_{\rm NL}^{\rm infl,GR}\right]
\label{eq:fnlGRand2ndorder}
\end{equation} 
where
\begin{equation}
f_{\rm NL}^{\rm infl,GR}=-\frac{5}{3}\left[1-\frac{5}{2}\frac{{\bf k}_1 \cdot {\bf k}_2}{k^2}\right].
\label{eq:fnlGR}
\end{equation}

This expression contains two contributions: the first concerns the primordial NG and the second ($f_{\rm NL}^{\rm infl,GR}$) is due to the horizon-scale PN corrections.
This contribution to non-Gaussianity has recently received attention. In particular,  for large-scale modes that enter the horizon deep in matter dominated era, no perturbation growth is expected before matter domination so their entire growth history can be modelled in the standard way assuming matter domination. Non-Gaussianity introduced by non-linear growth is negligible on very large, linear scales. Still at very large scales comparable to the Hubble radius  there is a contribution, $f_{\rm NL}^{\rm infl,GR}$, to NG arising from  GR corrections. This was first pointed out in \cite{BMR2005}, but it was not until 2009 that this issue was reconsidered in light of evaluating the contribution to the observed $f_{\rm NL}$ of LSS, from nonlinear
growth of modes that entered the horizon during the radiation era \cite{Fitzpatrick} and in light of making a  possible connection to observable quantities \cite{Yoo2009,wands&slosar}. In a pioneering work, the authors of \cite{Yoo2009} pointed out that in a general relativistic description of galaxy clustering one automatically includes several effects which were perviously dealt with separately; indeed the observed redshift and position of galaxies are affected by matter fluctuations and gravitational waves between the galaxies and the observer, and the volume element constructed by using the GR observables differs from the physical volume occupied by the observed galaxies. 
On the other hand the authors of \cite{wands&slosar} argued that  the comoving time-orthogonal gauge is the correct reference to obtain predictions for the LSS power spectrum on large scales.

Other approaches to obtain $f_{\rm NL}^{\rm infl, GR}$ have since been presented in the literature see e.g., \cite{BHMW14}. Despite the apparent difference in the equations,  it is easy too see that our Eq.~\eqref{eq:fnlGR} and equation (A.12) of \cite{BHMW14} become  exactly the same in the squeezed limit where NG is maximal and  on large scales.

The non-Gaussianity arising from Eqs.~(\ref{eq:fnlGRand2ndorder},\ref{eq:fnlGR}) therefore has the following characteristics: 
\begin{itemize}
\item Inflationary models different from the standard slow roll would yield corrections to the first term. In particular in models where fluctuations are created by an additional light field different from the inflation would produce a large local non Gaussianity.  Models with higher derivatives operators or modifications of the vacuum state would also produce extra contributions, not described by $a_{\rm nl}-1$.

\item $f_{\rm NL}$ is non-zero even if  the --strictly speaking-- primordial contribution is zero because of the presence of $f_{\rm NL}^{\rm infl,GR}$.
\item  $f_{\rm NL}^{\rm infl,GR}$ is peculiar to inflationary initial conditions. Perturbations on super horizon scales are needed to initially feed the GR correction terms. The significance of this term is analogous to the well-known large-scale anti-correlation between CMB temperature and E-mode polarization:  it is a consequence of the properties of the inflationary mechanism to lay down the primordial perturbations. In standard slow roll inflation $f_{\rm NL}^{\rm infl,GR}> |a_{\rm NL} -1|$, meaning that GR corrections dominate $f_{\rm NL}$. 
\end{itemize}
 
It is  clear that a detection or measurement of  $f_{\rm NL}^{\rm infl,GR}$ would have profound implications for our understanding of the behaviour of GR on large scales and it would offer a powerful consistency check for  one of the basic tenets of modern cosmology. 
 
\section{Detectability of GR effects on large-scales  halo bias}
It has been shown that primordial non-Gaussianity affects the clustering of dark matter halos on large, linear scales, inducing a scale-dependent bias. This scale dependence is particularly marked and grows at large scales for the case of local non-Gaussianity \cite{MV08,DDHS08}.  This can be physically understood by considering that halo clustering can be modeled by that of  regions where the (smoothed) linear dark matter density field exceeds a suitable threshold. For massive halos, the threshold is high compared to the field {\it rms}. The two-point correlation function of regions  above a high threshold for a general non-Gaussian field has an analytical expression \cite{MLB86} which depends on all higher-order (connected) correlations but for most cases in cosmology only terms up to the  three-point correlation function (the bispectrum) matter.
Thus, on large scales,   there is a correction to the standard halo bias due to the presence of primordial non-Gaussianity which can be written as:
\begin{equation}\label{eq:dboverb}
\frac{\Delta b_h}{b_h}= \frac{\Delta_c}{D_+(z)}\beta^{\rm NG}(k)\,.
\end{equation}
where $\Delta_c$ denotes the  density threshold for halo collapse.
It is clear that a scale-dependent $\beta^{\rm NG}(k)$ could in principle give a distinctive detectable signature on the observed (tracers) power spectrum. 

\begin{figure}
  \centering
\includegraphics[trim= 30 200 0 210,clip,width=10cm]{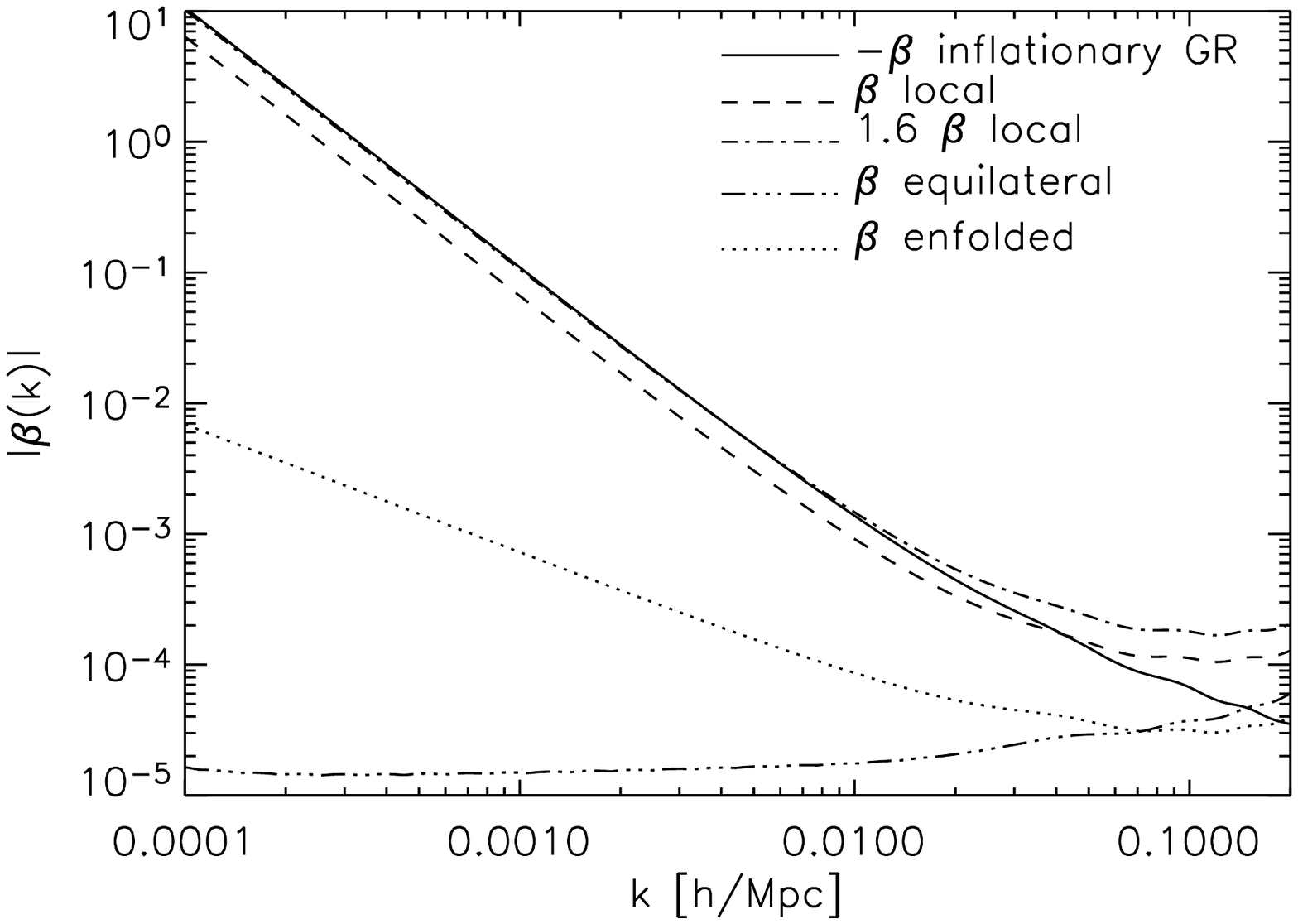}
\caption{
The scale dependence of the large-scale halo bias induced by a non-zero bispectrum. The
solid line shows the absolute value of $\beta^{NG}$ for the inflationary, GR
correction large-scale structure bispectrum. Note that the quantity is actually negative. The dashed line shows $\beta^{NG}$ for the local
type of primordial non-Gaussianity for $f_{\rm NL}=1$; it is clear that the scale-dependent bias effect due to the inflationary bispectrum mimics a local primordial non-Gaussianity.  The dot-dot-dot-dashed line shows the effect of
equilateral non-Gaussianity  and the dotted line shows the enfolded type both for  $f_{\rm NL} = 1$.  Figure reproduced with permission from \cite{VM09}.}
\label{fig:nghalobias}
\end{figure}

Biasing, a small-scale phenomenon, can affect the power spectrum on very large scales if the non-Gaussianity induce strong mode-coupling between small and large-scale modes. For this reason,  local and inflationary non-Gaussianity leave a strongly scale-dependent signature on the halo clustering, but other non-gaussianities such as equilateral and enfolded type have a much smaller effect \cite{VM09}. This can be appreciated in Fig.~(\ref{fig:nghalobias})  where we plot the magnitude of the scale-dependent factor $\beta^{\rm NG}$ for several choices of non-Gaussianities. 

In other words, halo bias is sensitive to bispectrum  configurations in which one of the three Fourier modes is much smaller than the other two, the so-called squeezed limit. In the local case the  bispectrum in squeezed limit goes like $1/k^3$ while other models of primordial non-Gaussianity have a less strong k dependence (e.g., $1/k$). In the squeezed limit the  GR correction  $f^{\rm infl. GR}_{\rm NL}$ scales like local non-Gaussianity, making therefore the effect potentially accessible by observations. 

It has been shown \cite{VM09} that  at the effects of the halo bias $f^{\rm infl. GR}_{\rm NL}\simeq -1.6$ and it behaves as a local non-Gaussianity. This is not unsurprising: in the squeezed limit $f_{\rm NL}^{\rm infl.GR}\sim -5/3$ (see Eq. \ref{eq:fnlGR}) with a sub-dominant additive (negative) correction that goes like  $1/k^2$ (i.e., like the orthogonal template in this limit, which is a superposition of the equilateral and enfolded  ones). In fact, due to this correction we see  --Fig.(\ref{fig:nghalobias})--- that at $k>0.01\,h/$Mpc the effect of $f_{\rm NL}^{\rm infl, GR}$ starts to deviate from the local behaviour.   An identical result is obtained when considering the $f^{\rm infl, GR}_{\rm NL}$ obtained in the Poisson gauge \cite{Pillepich} (once the constant, pure gauge, modes appearing in the Poisson-gauge expression are ignored), which is encouraging  as we recover that measurable quantities are indeed gauge-independent.

Note that since it is the squeezed limit that matters for the halo bias, the fact that different approaches to compute $f_{\rm NL}^{\rm infl,GR}$ coincide in this limit, highlights the robustness of the prediction.

Thus forecasts and constraints on $f_{\rm NL}$ for the local case, obtained from the shape of the large-scale halo power spectrum, can be suitably re-interpreted for   $f^{\rm infl, GR}_{\rm NL}$.
Since the scale-dependent, non-Gaussian halo bias effect was presented in 2008,  there have been several works on using the effect to constrain local NG from available data and on forecasting the performance of the approach from future data sets.
Error bars on $f_{\rm NL}$ obtained from current data range from optimistic $\pm 20$ e.g.,\cite{Slosarcurrent, Xiacurrent, Giannantonio2014} to the conservative $\pm 40$ \cite{Leistedt}, far too large to  be relevant for  detecting $f^{\rm infl, GR}_{\rm NL}$. However, forecasted error bars range form the more conservative $\pm$ few (e.g., \cite{CVM08,Cunha,Fedeli, Giannantonio2012}) to the optimistic $\pm 0.1$ (e.g.,  \cite{Hamaus,Yamauchi}). This would indeed guarantee a highly significant detection of the GR effect, if all systematic effects were under control (for examples of possibly systematic effects acting both in giving false positive or false negative see \cite{Reidassembly, PorcianiRoth,Leistedt}). It is encouraging that the forecasted error-bars do not depend on the underlying assumptions about the cosmological model because  the scale-dependent halo bias shows minimal covariance with cosmological parameters including non-standard $\Lambda$CDM  parameters such as dark energy evolution (e.g., \cite{Carbonedegen, Giannantonio2012}). The smallest  forecasted error-bars are obtained employing the multi-tracers technique proposed in Ref.~\cite{SeljakCV}. In fact the measurement of the power spectrum shape on large scales is dominated by cosmic variance because of the finite number of modes in the surveyed volume. While normally cosmic variance imposes an error-floor on measurable quantities, here this can be circumvented by comparing the density field of tracers with different bias. In this way, the relative bias of two (or more) tracers is not affected by cosmic variance, it is scale independent in the absence of NG, but picks up a scale dependence in its presence. If on large scales tracers are linearly biased but not  stochastic, the error on $f_{\rm NL}$ is only limited by noise. More sophisticated techniques have been further developed using several different tracers and optimally weighting them to minimise the statistical error on $f_{\rm NL}$ e.g.,~\cite{Hamaus} and refs. therein. In the wide-field surveys era,   a large faction of the full sky will be surveyed over several wavelengths thus mapping widely differently biased tracers and making this approach  feasible.  Compared to the standard approach, the multi-tracers one can reduce the error-bars on $f_{\rm NL}$ by a factor $\sim 10$.

This opens up the possibility to clearly  detect the horizon-scale GR signal from the next generation cosmological surveys.

\newpage
\section{Towards GR corrections on all scales: a PN approach}

In Sec.~\ref{NGlarge} we have seen that even on large scales, quantities computed beyond the linear approximations are needed to describe some observable effects. And we have seen that there is 
a complex interplay between sources non-linearities due to evolution, due  to the fact that the observable quantity is related in a non-linear way to the theoretically modelled quantity, due to the breakdown of the Newtonian 
approximation  and finally to the initial set up of perturbations. 
A unified description able to deal with all these effects and work at all scales would be highly valuable. Here we describe an alternative route which has been recently analysed: the Post-Newtonian approximation of cosmological perturbations.

Going beyond the Newtonian approximation and standard perturbation theory is certainly an ambitious task and cannot be accomplished without some approximation scheme. A description of the fully non-linear dynamics of the perturbations in the framework of GR can be carried out by means of effective fluid description for small-scale non-linearities, as proposed in \cite{baumann} or by means of an effective field theory approach, as in \cite{porto}. An elegant framework to describe nearly FRW space-time but where the deviations from the homogeneous density are not assumed to be small was instead proposed in \cite{wald1} and it is used to compare Newtonian and GR cosmology in \cite{wald2}.
Also, exact analytical solutions of Einstein's equations, can be important tools to investigate non-linear effects, e.g., in studying the path of photons throughout inhomogeneities, 
although in simplified contexts or in the presence of some symmetry in the problem, see e.g.,~\cite{meures1} and \cite{meures2}.

Alternatively, we should seek for a relativistic and non-perturbative approach, capable to disentangle the Newtonian from the relativistic contributions. The PN approximation of GR could be the key ingredient for this purpose: it provides the first relativistic corrections for a system of slowly moving particles bound together by gravitational forces and thus it can be used to account for the moderately non-linear gravitational field generated during the highly non-linear stage of the evolution of matter fluctuations on intermediate scales. It is a crucial improvement of both the aforementioned approximations, as it could bridge the gap between relativistic perturbation theory and Newtonian structure formation, providing a unified approximation scheme able to describe the evolution of cosmic inhomogeneities from the largest observable scales to small ones, including also the intermediate range, where the relativistic effects cannot be ignored and non-linearity starts to be relevant. The key novelty is that the PN approximation has by construction a direct correspondence with Newtonian quantities: the PN expressions are sourced only by the non-linear Newtonian terms which can be extracted e.g.,~from N-body simulations. Such a correspondence becomes increasingly important especially when studying frame-dependent quantities. Since we deal with the perturbations with respect to a Newtonian background, it is preferable to choose gauges with a clear Newtonian interpretation. The most suitable are therefore the Poisson gauge and the synchronous and comoving gauge, which reduce to the Eulerian and Lagrangian picture for the fluid dynamics in Newtonian gravity, respectively: see \cite{VMMN} for the fully non-perturbative comparison and \cite{rampf2} for a recent discussion of the second-order GR gauge transformation developed in \cite{MMB}.
In the Poisson gauge the metric perturbations remain small, even when the density locally blows up at particle orbit-crossing (this is because Green's functions act as a smoothing kernel) and there are not gauge ambiguities. It is often used to calculate the photon path but it is not directly related to halo bias. On the contrary, the synchronous and comoving gauge is suitable for halo bias calculations but it is not well-defined beyond first orbit crossing. 

It is also worth recalling that perturbation theory in the Lagrangian picture is more powerful than the Eulerian one: one searches for solutions of perturbed trajectories about the initial position of fluid element instead of the perturbing density and velocity fields, as in the Eulerian approach. The important point is that a slight perturbation of the Lagrangian particle paths carries a large amount of non-linear information about the corresponding Eulerian evolved observables, since the Lagrangian picture is intrinsically non-linear in the density field. In Newtonian gravity both Eulerian and Lagrangian perturbation theory were widely studied in the past: see e.g.,~\cite{buchert89}-\cite{catelan95E} and refs. therein. For a clear comparison between Eulerian and Lagrangian perturbation theory and the Zel'dovich approximation see \cite{bouchet}.

Obtaining the metric in the PN approximation in the synchronous and comoving gauge is extremely challenging, mostly because the Newtonian background metric, yet describing non-linearities, is space- and time-dependent, and the expression for the PN perturbation in the spatial metric, i.e. the solution of the PN Einstein equations, is it not easy to obtain for all modes, see \cite{mater}. This very fact has so far prevented from proceeding in this direction because of the computational complexity, except for symmetric configurations, as the one we consider in section \ref{planep}. Another approach is to consider the PN approximation in the Poisson gauge, which is computationally simpler, e.g., because the conformal spatial background metric is the Euclidean metric, and transforms to the synchronous and comoving gauge in a non-perturbative way. This is the approach adopted in \cite{VMMN} and \cite{VMMPN}. Although it may appear complicated, it is rather very helpful, already in the Newtonian limit: most derivations of the Newtonian limit of GR are coordinate-dependent, thus a precise understanding of the Newtonian correspondence, e.g., between the Eulerian and the Lagrangian frame, has to be considered as the starting point for studying the gauge dependence when we want to add GR corrections in a perturbed space-time from a non-perturbative perspective. \\

\subsection{The Newtonian limit}
The Newtonian limit in the Poisson gauge, defined in \cite{bert96}, is given by \footnote{Indices notation: we use $A, B, ...$ for spatial Eulerian indices, $\a, \b, ...$ for spatial Lagrangian and $a, b, ...$ to indicate space-time indices in any gauge.}
\begin{equation}\label{newtE}
ds^2=a^2\left[-\left(1+2\frac{\varphi_g}{c^2}\right)c^2 d\eta^2 +\delta_{AB}dx^A dx^B\right],
\end{equation}
where $\varphi_g$ is the Newtonian gravitational potential. The point of view illustrated above is purely Newtonian: the matter is viewed as responsive to the gravitational field given by $\varphi_g$ and the Newtonian order in the metric is established considering just the equations of motion for the fluid, the Euler and continuity equation, i.e. the lowest order in the $1/c^2$ expansion of $\nabla^a T_{aK}=0$ and $\nabla^a T _{a0}=0$, where only the time-time component of the metric is required, with $\varphi_g$ satisfying the cosmological Poisson equation
\begin{equation}
\nabla_{x}^2\varphi_g=4\pi G a^2 \rho_b \delta
\end{equation}
given by the time-time component of the Einstein equations, where $\rho_b$ is the FRW background matter density, $\delta=\left(\rho-\rho_b\right)/\rho_b$ is the density contrast and $a(t)$ is the 
scale-factor, which obeys the Friedmann equation. This implies that for a fluctuation of proper scale $\lambda$ we have
\begin{equation}
\frac{\varphi_g}{c^2} \sim \delta\left(\frac{\lambda}{r_H}\right)^2 \;,
\end{equation}
where $r_H=cH^{-1}$ is the Hubble radius. This very fact tells us that the weak-field approximation does not necessarily imply small density fluctuations, 
rather it depends on the ratio of the perturbation scale $\lambda$ to the Hubble radius.
That is why Newtonian gravity is widely used to study structure formation at small scales, also in the non-linear regime. However, it is well-known that it fails to produce an accurate description of photon trajectories: it is well know that the Newtonian estimate of the Rees-Sciama effect and of gravitational lensing using the line element \eqref{newtE} is incorrect by a factor of two. This is because when considering equation \eqref{newtE} as the Newtonian limit we only account for the matter fluctuations, described by the gravitational potential, and neglect the spatial curvature (the space is flat in Newtonian gravity) which does affect the path of photons. The correct calculation involves the so-called weak-field limit of GR, which is valid for slow motions of the sources of the gravitational field, but allows test particles to be relativistic. The related metric is perturbed also in the space-space component, allowing only the PN scalar $-2\varphi_g/c^2$, given by the trace part of the space-space components of Einstein equations, which is the lowest PN contribution to the spatial curvature. The weak-field approximation reads
\footnote{We remark that the line element in eq. \eqref{weakE} is not referred to the so-called longitudinal gauge, where vector and tensor modes are set to zero by hand at all orders and only the scalar mode in the spatial metric is present. Strictly speaking, it is not even a gauge, since only one among the six physical degrees of freedom in the metric are allowed.} 
\begin{equation}\label{weakE}
ds^2=a^2\left[-\left(1+2\frac{\varphi_g}{c^2}\right)c^2 d\eta^2 +\left(1-2\frac{\varphi_g}{c^2}\right)\delta_{AB}dx^A dx^B\right].
\end{equation}

In the synchronous and comoving gauge the Newtonian dynamics is described in terms of the mapping $\mathbf{x}(\mathbf{q},\tau)=\mathbf{q}+\mathbf{\mathcal{S}}(\mathbf{q},\tau)$ between the Eulerian, i.e. evolved, position and the Lagrangian, i.e. initial, position of fluid particles. The Newtonian limit in \cite{mater} is found by writing the spatial conformal metric tensor as $\gamma_{\a\b}=\d_{AB}\mathcal{J}^{A}_{\a}\mathcal{J}^{B}_{\b} $ where $\mathcal{J}^{A}_{\a}$ is the Jacobian matrix of the transformation of spatial coordinates. As in the Poisson gauge, in this metric we find just what we need for the Newtonian Lagrangian equations of motion, namely the Raychaudhuri equation, the continuity equation and the momentum constraint.\footnote{Consistently, the Raychaudhuri equation transforms in the Eulerian frame in the Euler equation and the momentum constraint in Lagrangian space is the irrotationality condition for the dust. Finally, the continuity equation is solved exactly in the Lagrangian frame, see \cite{mater, VMMN}.}

\subsection{Towards the PN approximation in the synchronous and comoving gauge}

The authors of \cite{VMMN} show that, from a GR perspective, writing the spatial metric in the synchronous and comoving gauge as $\gamma_{\a\b}=\d_{AB}\mathcal{J}^{A}_{\a}\mathcal{J}^{B}_{\b} $ leads to inconsistencies:
e.g., the four-dimensional (conformal) curvature not being preserved by the spatial transformation given above.
This is not surprising: at lowest order in the $1/c^2$ expansion in the Poisson gauge only the perturbation of the time-time component of the metric contributes to the scalar curvature $\mathcal{R}$, whereas in the synchronous and comoving gauge 
we need also one PN mode (the scalar $\chi$, see below) in the spatial PN metric. Indeed the four-dimensional (conformal) curvature at the lowest order in the $1/c^2$ expansion reads,
\begin{equation}
\mathcal{R}=2\vartheta'+\vartheta^2+\vartheta^\mu_\nu\vartheta^\nu_\mu+\,^{(3)}\!{\cal R}^{PN}\,.
\end{equation}
where $\vartheta^\mu_\nu$ is the Newtonian peculiar velocity-gradient tensor and $^{(3)}\!{\cal R}^{PN}=-2\mathcal{D}^2\chi$,\footnote{ $\mathcal{D}^2=\mathcal{D}^\ss\mathcal{D}_\ss$, $\mathcal{D}_\ss$ being the covariant spatial derivative in the Newtonian limit, is the spatial PN curvature.}

In order for the scalar $\mathcal{R}$ to be conserved at the lowest order, we have to start from the weak-field metric in the Poisson gauge 
\begin{equation}\label{weakE}
ds^2=a^2\left[-\left(1+2\frac{\varphi_g}{c^2}\right)c^2 d\eta^2 +\left(1-2\frac{\varphi_g}{c^2}\right)\delta_{AB}dx^A dx^B\right].
\end{equation}
and consider the gauge transformation to the higher order, including the effect of the time transformation, written as 
\begin{equation}\label{timetrans}
\tau=\eta-\frac{1}{c^2}\xi^0(x^A,\eta),
\end{equation}
keeping only the scalar part in the resulting PN transformed spatial metric. The PN weak-field limit, becomes in the synchronous and comoving gauge, see \cite{VMMN},
\begin{equation}
ds^2=a^2\left[-c^2 d\tau^2 +\left(1+\frac{\chi}{c^2}\right)\mathcal{J}^{A}_{\a}\mathcal{J}^{B}_{\b}\d_{AB}dq^\a dq^\b\right],
\end{equation}
where
\begin{equation}
\chi= 2\mathcal{H}\xi^0_\mathcal{L} -2\varphi_g^\mathcal{L}-\Upsilon^\mathcal{L}.
\end{equation}
In the latter expression, the potential $\Upsilon$ is given by
\begin{equation}
\mathcal{D}^2\Upsilon=-\frac{1}{2}\left(\vartheta^2-\vartheta^\mu_\nu\vartheta_\mu^\nu\right)
\end{equation}
and for the time transformation we have\footnote{The constant of integration $C(q^\a)$ in equation (4.35) in \cite{VMMN} represents the gauge freedom of the synchronous and comoving gauge and can be set to zero.}

\begin{equation} \label{xi0solutionconcl}
\xi^0=\frac{1}{a}\int^\eta_{\eta_{in}}a\left(-\varphi_g+\frac{1}{2}v^Av^B\delta_{AB}\right) d\tilde{\eta}
\end{equation}
and
\begin{equation}
v^{K}=\frac{\partial \xi^0_\mathcal{L}}{\partial q^\lambda}\mathcal{J}_F^\lambda\delta^{FK} ,
\end{equation}
being $\varphi_g$ the gravitational potential and $v^A$ the Eulerian peculiar velocity.

Starting again from the weak-field metric and transforming the time according to \eqref{timetrans}, keeping also the tensor part resulting from the time transformation, we are able to obtain also the (trace- and divergence-less) PN tensor modes $\pi_{\a\b}$ of the spatial metric
\begin{equation}\label{WeakL}
ds^2=a^2\left[-c^2 d\tau^2 +\left(1+\frac{\chi}{c^2}\right)\mathcal{J}^{A}_{\a}\mathcal{J}^{B}_{\b}\d_{AB}+\frac{1}{c^2}\pi_{\a\b} dq^\a dq^\b\right],
\end{equation}
where
\begin{equation}\label{pi}
\mathcal{D}^2 \pi_{\a\b}=\mathcal{D}_\a\mathcal{D}_\b\Upsilon+\mathcal{D}^2\Upsilon\gamma_{\a\b}+2\left(\vartheta\vartheta_{\a\b}-\vartheta_{\a\mu}\vartheta^\mu_\b\right)\,.
\end{equation}
This metric in equation \eqref{WeakL} represents the first GR correction to the Newtonian metric in the synchronous and comoving gauge, in the sense that the PN scalar field $\chi$ and the PN tensors $\pi_{\a\b}$ are required at the lowest order in the energy constraint and in the evolution equation respectively, whereas only the Jacobian spatial matrix $\mathcal{J}^{A}_{\a}$ is required for the (Newtonian) Raychaudhuri equation and momentum constraint, see \cite{mater}. In other words, the metric \eqref{WeakL} solve the complete set of the Einstein equations, at the lowest order. 
The appearance of the tensor modes in the synchronous and comoving gauge at the PN order, whereas in the Poisson gauge they are $\mathcal{O}(1/c^4)$ is once again due to the gauge dependence of the Newtonian approximation. The Lagrangian dynamical variable is the peculiar velocity-gradient tensor, which in this gauge is given by the time derivative of the spatial metric $\vartheta_{\a\b}=1/2\gamma'_{\a\b}=\mathcal{J}_{\a B}\mathcal{J}^{B'}_{\b}$, where the second equality holds in the Newtonian limit. This is the source in the equations for the PN spatial Ricci tensor $^{(3)}\!{\cal R}^{\a\,PN}_{\b}=\mathcal{D}^\a\mathcal{D}_\b\chi-\mathcal{D}^2\chi\delta^\a_\b+\mathcal{D}^2\pi^\a_\b$, namely the energy constraint for $\chi$
\begin{equation}
\vartheta^2 - \vartheta^\mu_{~\nu} \vartheta^\nu_{~\mu} + 4 \mathcal{H}
\vartheta -16 \pi G a^2\rho_b \delta  =  2\mathcal{D}^2\chi
\end{equation}
and the evolution equations \eqref{pi}.

In order to find the complete PN metric in synchronous and comoving gauge, where one more scalar and the vector modes are needed, we have to start from the fully PN approximation in the Poisson gauge \eqref{PNE}, i.e. including a divergence-less vector contribution in the time-space component and the higher-order scalar in the time-time component:
\begin{equation} \label{PNE}
ds^2=a^2\left[-\left(1+\frac{2\varphi_g}{c^2}+\frac{\Phi}{c^4}\right) c^2 d\eta^2 +\frac{\omega_A}{c^3}c 
d\eta dx^A+\left(1-\frac{2\varphi_g}{c^2}\right)\delta_{AB} dx^Adx^B\right]
\end{equation}
and consider the higher orders in the transformation of the spatial coordinates and of the time. 

\subsection{PN plane parallel dynamics} \label{planep}

We specialise the PN Lagrangian dynamics in the synchronous and comoving gauge to globally plane-parallel configuration, i.e. to the case where the initial perturbation field depends on a single spatial coordinate\footnote{In this section we also assume a vanishing cosmological constant}. The leading order of our expansion, corresponding to the Newtonian background, is the Zel'dovich approximation \cite{zel}, which, for plane-parallel perturbations in the Newtonian limit, represents an exact solution. The peculiarity of this treatment, at any order, is that, while the displacement vector is calculated from the equations at the required perturbative order, all the other dynamical variables, such as the mass density, are calculated exactly from their non-perturbative definition. Since the equations in Lagrangian coordinates are intrinsically non-linear in the density, what comes out is a fully non-linear description of the system, which, though not being generally exact, mimics the true non-linear behaviour; see also \cite{Tassev} for a re-analysis of the Zel'dovich approximation in connection to the mildly non-linear range of scales probed by baryonic acoustic oscillations in the matter power spectrum. The Zel'dovich approximation arises in Newtonian theory and one might ask how to correctly extend it to GR. This problem has already been discussed a number of times, often reducing to a standard second-order treatment in the synchronous and comoving gauge, 
as e.g., in \cite{russ}, thereby partially missing the non-perturbative character of the Zel'dovich approximation. On the contrary in the spirit of the PN approximation explained above, our approach aims at obtaining a non-perturbative description of both 
metric and fluid properties (velocity-gradient tensor and mass density), within the PN approximation of GR: our expansion in inverse powers of the speed of light is fully non-perturbative from the point of view of standard perturbation theory; thus our results contain all second and higher-order terms of standard perturbation theory calculations, as long as they are PN and one deals with the plane-parallel dynamics. In our approximation scheme the Zel'dovich solution represents the Newtonian background over which PN corrections can be computed as small corrections.

The high symmetry of the plane-parallel configuration allows us solved the PN Einstein equations, with linear and second-order initial conditions motivated by inflation. The PN Zel'dovich solution reads, \cite{VMM1}:
 \begin{eqnarray} \label{soluzione1}
\gamma_{11}&= &\left(1-\frac{\tau^{2}}{6}\du^{2}\varphi\right)^{2}+\frac{1}{c^{2}}\left\{ \left[\frac{5}{108}
\tau^2\left(\left(4\left(a_{\rm nl}-1\right)-1\right)\left(\du\varphi\right)^2+\right. \right.\right. \nonumber \\
\hphantom{\times\biggl\{}&+& \left. \left.\left.  \left(4\left(a_{\rm nl}-1\right)-4\right)
\varphi\du^{2}\varphi\right)+\frac{5}{576}\tau^4\du^{2}\varphi\left(\du\varphi\right)^2 
\right]\left(6-\tau^2\du^{2}\varphi\right)-\frac{5}{54}\varphi\left(6-\tau^2\du^{2}\varphi\right)^2\right\} \nonumber \\
\gamma_{22}&=& 1+\frac{1}{c^{2}} \left(-\frac{10}{3} \varphi+\frac{5}{18}\tau^{2}(\du\varphi)^{2}\right) \\
\gamma_{33}&=& 1+\frac{1}{c^{2}} \left(-\frac{10}{3} \varphi+\frac{5}{18}\tau^{2}(\du\varphi)^{2}\right). \nonumber
\end{eqnarray}  
This expression describes the plane parallel dynamics at all scales of interest: at smaller scales the (exact) Zel'dovich Newtonian term is dominant, at intermediate scales it is taken into account the PN contribution, including the primordial NG. Once expanded in standard perturbation theory, our PN metric \eqref{soluzione1} reproduces the correct GR description up to second order at largest scales for one-dimensional perturbations.
This solution is valid as long as we restrict our analysis to suitably large scales, where the cosmological dynamics is only governed by gravitational self-interaction of the dark matter. At smaller scales the irrotational and pressureless fluid approximation breaks down and shell-crossing singularities appear as an artefact of the extrapolation of this approximation beyond the point at which velocity dispersion has become important. Let us just remark an important result of this analysis of the final stages of plane-parallel collapse, a pancake-like singularity, related to the cosmological backreaction. This (very controversial) topic is far outside the subject of the present paper; we refer to \cite{backrew} for a review. An important result is, however, that caustic formation does not affect the averaged quantities: both the spatial curvature and the matter density diverge but the mean quantities keep perfectly smooth; see \cite{VMM1} for the explicit expressions. 
This shows that caustics do not lead to any instability in the averaged quantities.

\section{Concluding remarks}
\label{sec:conclusions}

In this paper we have discussed some key examples of how a fully GR perspective of cosmology may affect our interpretation of cosmological observables and modify the theoretical approach to LSS evolution.

In particular, in section \ref{NGlarge} we analysed the NG on large scales following \cite{BMPR2010}. Let us compare our approach with that of \cite{BHMW14}\footnote{See also \cite{BHW14} for the same approach at all perturbative orders.}, limiting to their results on large scales, and \cite{DDHS08}. 
In \cite{BMPR2010} the authors relate the GR density contrast to an effective potential by means of the kernel of Eq.~\eqref{kddthth}. 
We note that, considering local NG in the squeezed limit, which is the dominant configuration for halo bias, our equation \eqref{eq:fnlGR} and equation (A.12) of \cite{BHMW14}, obtained by simply Fourier transforming the density field, give exactly the same contribution at large scales, i.e., $f_{\rm NL}^{\rm infl.GR}\sim -5/3$ with a correction that goes like  $1/k^2$. The same result is obtained following \cite{DDHS08}, where the analytical approximate estimation holds for local NG only. See also \cite{schmidt} for an alternative approach in the squeezed limit.

In most cosmological applications either Newtonian gravity is  assumed (at small scales and late-times where non-linearities are important) or  (quasi-linear) relativistic perturbation theory around a homogeneous background, at large-scales and early times. However,  there are specific instances where these two approximations are not sufficient. We have illustrated two of such cases: the study of large-scale structures to horizon-size scales and the analysis of non-linear post-Newtonian effects in the Lagrangian frame, which is a key ingredient e.g., to analyse the back-reaction of perturbations on the average universe expansion rate (see, e.g.,~\cite{ehlers,KMR, backrew}) or the trajectory of relativistic particles in the gravitational field generated by cold dark matter perturbations.
Horizon-size GR effects induce a correction $f^{\rm infl, GR}_{\rm NL}$ to the primordial NG parameter, which governs the amplitude of the gravitational potential bispectrum. The shape and amplitude of this bispectrum contribution is peculiar in several aspects: {\it i)} Perturbations on super-Hubble scales are needed in order to initially feed the GR correction terms. In this respect, the significance of this term is analogous to the well-known large-scale anti-correlation between CMB temperature and E-mode polarization: it is a consequence of the properties of the inflationary mechanism to lay down the primordial perturbations. {\it ii)} Standard slow-roll single field inflation implies that the second-order comoving curvature perturbation is small, and thus the  primordial contribution to $f_{\rm NL}$ is subdominant compared to  $f^{\rm infl, GR}_{\rm NL}$. Consequently $f_{\rm NL}$ is non-zero (i.e., on the LSS there is an effect akin to primordial non-Gaussianity), even if inflationary non-Gaussianity were to be exactly zero ($|a_{\rm nl}-1|=0$).  {\it iii)} The corresponding bispectrum is very close to the local shape especially in the squeezed configuration (where the local shape is maximal) and in this limit has a well defined amplitude $f^{\rm infl, GR}_{\rm NL}=-5/3$.

The squeezed limit of the bispectrum can be accessed  directly  via the so-called scale-dependent halo bias: in the absence of non-Gaussianity ($f_{\rm NL}=0$),  different dark matter tracers, who populate density peaks above a threshold, on large scales have a scale independent bias. This bias picks up a marked scale dependence  if  $f_{\rm NL}\ne 0$.  Thus results from  any analyses done for the local type of non-Gaussianity can be re-interpreted in terms of the GR effects of interest here.

While current surveys do not have enough statistical power to detect  $f^{\rm infl, GR}_{\rm NL}$, forecasted  error-bars from future surveys guarantee a high significance detection and tight measurement of this quantity. Of course, this holds provided that all systematic effects (both in the data and in the theoretical modelling)  are unimportant or can be kept under control.
Recent work \cite{Baldauf} has suggested that even better signal-to-noise for NG could be obtained by considering the halo bispectrum at smaller, mildly non-linear scales, rather than the power spectrum at large, linear scales. Of course, because of the separation of scales the two statistics could be straightforwardly  combined, dramatically shrinking  the  error-bars on $f_{\rm NL}$.  Some considerations however curb one's immediate enthusiasm in translating this into a dramatic improvement in the signal-to-noise for  $f^{\rm infl, GR}_{\rm NL}$. 
On mildly non-linear scales and away from the squeezed limit the  bispectrum of GR effects non-Gaussianity  is  not strictly local.  Still this feature could be used for distinguishing  $f^{\rm infl, GR}_{\rm NL}$ from e.g., NG arising from multi-field inflation models. Most importantly however, bispectrum forecasts have been performed under heavily idealised conditions; for example, if bias (in the absence of non-Gaussianity) can be considered, linear, local and deterministic on very large scales, it is a drastically  more complex process on mildly non-linear scales (see e.g., \cite{HGMBoss} for a painful, first-hand experience with bispectrum and tracers bias from actual data).

Despite this cautionary note, there is  an interesting synergy between LSS and CMB that can be exploited to separate GR effects from  those of a possible NG primordial signal. If the simplest inflationary scenario holds,  from future LSS surveys the halo-bias approach is expected to detect a non-Gaussian signal very similar to the local type signal  which is due to large-scales GR corrections to the Poisson equation. This effect  leaves no imprint in the CMB  since $f^{\rm infl, GR}_{\rm NL}$ is there only in LSS. 
If  primordial non-Gaussianity is local with negative $f_{\rm NL}$ and CMB obtains a detection, then the halo bias approach should also give a high-significance detection (GR correction and primordial contributions add up), while if it is local but with positive $f_{\rm NL}$, the halo-bias approach could give a lower statistical significance  as the GR correction contribution has the opposite sign.  Once again the combination of these two observables can help enormously to test the basic tenets  of modern  cosmology.

A second direction of investigation discussed in this paper is the idea of going beyond the Newtonian approximation to study non-linear gravitational instability in cosmology. 
Recently PN-type approximations in the Poisson gauge have been considered e.g., in \cite{CM2005}, where the authors propose a hybrid approximation of Einstein field equations, which mixes PN and second-order perturbative techniques, 
and \cite{brunidrag, brunilensing} where an alternative approach to the Newtonian limit is proposed. The Newtonian and the PN approximation, in both the Poisson gauge and the synchronous and comoving gauge, deserve further investigation and the approach of \cite{VMMN} and \cite{VMMPN} is essential to deeply understand the connection between Newtonian gravity and GR in a cosmological context. 
Such results actually only represent the very first step towards an ambitious goal: finding a unified approximation scheme, from the linear to the non-linear scales, able to capture the most important GR effects on the LSS of the Universe. 

Recently some interesting and straightforward algorithms, able to capture some non-linear GR effects in numerical simulations of LSS, have been proposed: these techniques rely on suitable remapping from standard Newtonian N-body outputs (see e.g., \cite{baldauf, Chisari}.
One may imagine that the ultimate goal in GR cosmology should be that of performing numerical simulations of LSS formation using the full set of GR equations. 
In spite of the complication arising from the need of simultaneously solving several differential equations, one would 
have the tremendous conceptual advantage of dealing only with hyperbolic evolution equations (see e.g., \cite{ellis71}), hence restricting the need of (e.g., periodic) spatial boundary conditions to the construction of a self-consistent set of initial data 
(e.g., for the comoving curvature perturbation field) satisfying the relevant constraint equations on a suitable Cauchy hyper-surface.  
We envision that the next few years will be fruitful in these directions.
\subsection*{Acknowledgments}
EV thanks ``Fondazione Angelo della Riccia" for financial support.  LV's research is supported by the European Research Council under the European Community's Seventh Framework Programme FP7-IDEAS-Phys.LSS 240117 and in part by Mineco grant FPA2011-29678- C02-02. SM acknowledges partial financial support by the ASI/INAF Agreement 2014-024-R.0 for the Planck LFI Activity of Phase E2.

\end{document}